\documentclass{article}

\usepackage[portuguese]{babel}
\usepackage{xcolor}
\usepackage{arxiv}

\usepackage[utf8]{inputenc} 
\usepackage[T1]{fontenc}    
\usepackage{hyperref}       
\usepackage{url}            
\usepackage{booktabs}       
\usepackage{amsfonts}       
\usepackage{nicefrac}       
\usepackage{microtype}      
\usepackage{lipsum}		
\usepackage{graphicx}
\usepackage[authoryear]{natbib} 
\usepackage{doi}

\usepackage{amsmath}
\usepackage{amssymb}

\title{Cosmologia com aglomerados de galáxias}

\author{ \href{https://orcid.org/0000-0003-2652-0891}{\includegraphics[scale=0.06]{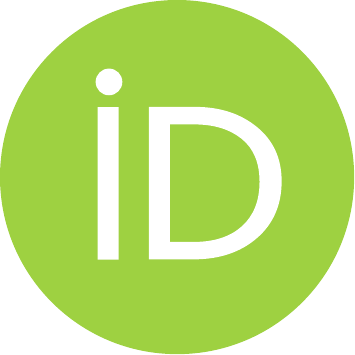}\hspace{1mm}M Penna-Lima}
		\\
	Instituto de Física, Universidade de Brasília, \\
	Campus Darcy Ribeiro, Brasília, 70.919-970, Brasil,\\
	\texttt{pennalima@unb.br} \\
}

\date{}

\renewcommand{\shorttitle}{Artigo} 

\begin{document}
\maketitle

\begin{abstract}
Os aglomerados de galáxias são importantes sondas cosmológicas, já que a abundância e a distribuição espacial desses objetos estão diretamente ligadas à formação de estruturas em grandes escalas. A maior fonte de incerteza nas restrições de parâmetros cosmológicos é originária das estimativas das massas dos aglomerados a partir da relação massa-observável. Além disso, os próximos grandes levantamentos fornecerão uma grande quantidade de dados, requerendo uma melhoria na precisão de outros elementos utilizados na construção das verossimilhanças de aglomerados. Portanto, uma modelagem precisa da relação massa-observável e diminuir o efeito dos diferentes erros sistemáticos são passos fundamentais para o sucesso da cosmologia com aglomerados. Neste trabalho, fazemos uma breve revisão da abundância de aglomerados de galáxias, e discussão de diferentes fontes de incerteza. 
\end{abstract}


\section{Introdução}
\label{sec:introducao}

 Os aglomerados de galáxias são as maiores estruturas ligadas do universo. A massa e o tamanho (diâmetro) típicos desses objetos variam entre $\sim 10^{13}$ - $10^{15} \, M_\odot$ e 1 - 10 Mpc, respectivamente. Historicamente, a concepção da existência dos aglomerados de galáxias ocorre somente no século XX, mas ela é baseada, entre outras, em observações de nebulosas cujos primeiros catálogos foram produzidos, independentemente, por Charles Messier e William Herschel no século XVIII. Em 1958 George O. Abell cria o primeiro catálogo de aglomerados, composto por 2,712 objetos ricos, massivos com \textit{redshift} até $z \lesssim 0,2$. Ampliando a cobertura de observação do céu, em 1989 são adicionados mais 1,361 aglomerados a esse catálogo~\citep{Abell1989}. Para mais detalhes sobre a evolução histórica nos primeiros estudos de aglomerados de galáxias, veja~\citep{Biviano2000}.

\citet{Zwicky1933} infere a presença de matéria escura no aglomerado Coma e, desde então, os aglomerados de galáxias têm sido utilizados no estudo da cosmologia~\citep{Allen2011}. Uma das grandes vantagens dos aglomerados é que eles emitem sinais em diferentes bandas do espectro eletromagnético como, por exemplo, raio-X, milimétrico, infravermelho, visível e rádio. A partir desses múltiplos sinais (além de efeitos gravitacionais) sabemos que os aglomerados, em geral, possuem a seguinte composição: $\sim 80 - 85 \%$ de matéria escura, $\sim 10 - 15\%$ de gás quente difuso/plasma intra-aglomerado (ICM)\footnote{Do inglês \textit{intracluster medium}.}, e $\sim 2\%$ de galáxias~\citep{Gastao2014}.

Grandes levantamentos que em breve entrarão em operação como o Vera C. Rubin {\it Legacy Survey of Space and Time} (LSST) \citep{lsst2012},  e o satélite {\it Euclid} \citep{Euclid2011}, por exemplo, transformarão a cosmologia observacional por produzirem um grande volume de dados. Esses dados diminuirão ainda mais a incerteza estatística nos ajustes de modelos cosmológicos e, portanto, as análises serão dominadas pelos erros sistemáticos\footnote{Os erros sistemáticos podem ser originados por falhas nos equipamentos/instrumentos e também na modelagem. Erros sistemáticos enviesam os resultados e esses não são atenuados com o aumento do número de medidas realizadas.}. Logo, este é um momento crucial para entender e modelar essas fontes de erro. Neste trabalho, revisaremos o uso de aglomerados de galáxias como sonda cosmológica dando foco à abundância de aglomerados para exemplificarmos e discutirmos algumas das fontes de incerteza. Para mais informações e detalhes no assunto, vejam, por exemplo, os artigos de revisão \citep{Allen2011} e \citep{Kravtsov2012}. 

Este trabalho está organizado da seguinte forma. Na Sec.~\ref{sec:contagem} apresentamos os elementos necessários para calcular a densidade do número de halos de matéria escura. A conexão com a contagem de aglomerados de galáxias é feita na Sec.~\ref{sec:proxy}. Discutimos, na Sec.~\ref{sec:likelihood}, algumas verossimilhanças que têm sido utilizadas na literatura para restringir parâmetros cosmológicos e/ou calibrar relações massa-observável. Finalmente, na Sec.~\ref{sec:incerteza}, apresentamos algumas fontes de erro que devem ser levadas em conta nas análises cosmológicas, assim como nas análises astrofísicas (calibração de massa, entre outros), e concluímos com as considerações finais (Sec.~\ref{sec:conclusao}). 

\section{Contagem de halos de matéria escura}
\label{sec:contagem}

A teoria de formação de estruturas fornece uma modelagem da instabilidade gravitacional em que descrevemos a matéria como um fluido do tipo poeira (sem colisões). Essa é uma boa aproximação para a matéria escura, já que ela interage essencialmente por gravitação, mas não é uma boa descrição para a matéria bariônica (principalmente quando atingi-se um regine não-linear). Neste contexto, descrevemos a formação dos chamados halos de matéria escura. 
O número médio de halos de matéria escura com redshift no intervalo $[z, z + dz]$ e massa $[M, M + dM]$ é  
\begin{equation}\label{eq:d2N}
\frac{\rm{d}^2N}{\rm{d}z \ \rm{d}\ln M} = A_{\text{survey}} \frac{c}{H(z)} D_A^2(z) \frac{\rm{d} n(M, z)} {\rm{d} \ln M},  \end{equation}
onde $A_{\text{survey}}$ é a área angular (ângulo sólido) do levantamento em rad${}^2$, $c$ é a velocidade da luz, $H(z)$ é a função de Hubble e $D_A(z)$ é a distância diâmetro-angular. A função de massa de halos, ou seja, a densidade número comóvel de halos com massa $M$ em $z$ é dada por
\begin{equation}\label{eq:mass_function}
 \frac{\rm{d} n(M, z)}{\rm{d}\ln M} = -\frac{\rho_m(z)}{M} f(\sigma_\text{\tiny{M}}, z) \frac{1}
			{\sigma_\text{\tiny{M}}} \frac{\rm{d} \sigma_\text{\tiny{M}}}{\rm{d} \ln M},  
\end{equation}
onde $\rho_m(z)$ é a densidade média de matéria do universo, $f(\sigma_\text{\tiny{M}}, z)$ é a função multiplicidade, e $\sigma_\text{\tiny{M}}$ é o desvio padrão do contraste de densidade de matéria filtrado em uma região esférica de raio $R$, a saber
\begin{equation}
   \sigma^2_\text{\tiny{M}} (R, z) = \int_0^\infty \frac{\rm{d}k}{2\pi^2} k^2 P(k, z)\vert W(k, R) \vert^2, 
\end{equation}
onde $P(k, z)$ é o espectro de potência linear de matéria, e $W(k, z)$ é a transformada de Fourier da função janela.
Em particular, consideramos uma função janela do tipo \textit{top-hat} que filtra uma região esférica de raio $R$ no espaço real,
\begin{equation}
  W(r, R) = \frac{3}{4\pi R^3}
    \begin{cases}
      1 & \text{para $r \leq R$}\\
      0 & \text{para $r > R$},
    \end{cases}       
\end{equation}
cuja transformada de Fourier é 
\begin{equation}
  W(k, R) = \frac{3}{(kR)^3} \left( \sin (kR) - (kR) \cos (kR) \right). 
\end{equation}
A massa do halo compreendida nesta região é, portanto, $M = \frac{4\pi}{3} \rho_m R^3$.\footnote{A massa do halo pode ser definida também em termos da densidade crítica do universo.} 

A partir da descrição acima, podemos identificar os diferentes termos que dependem explicitamente de modelo cosmológico, e que, consequentemente, fazem com que a abundância de halos de matéria escura seja uma sonda cosmológica. Esses termos são $H(z)$ (lembrando que $D_A(z)$ depende desta função), $\rho_m(z)$ e $P(k, z)$. 

Por sua vez, a função multiplicidade $f(\sigma_\text{\tiny{M}}, z)$, a qual concentra toda a informação do regime não-linear da formação dos halos, é quase-independente da cosmologia. Esse fato tem como fundamento o modelo do colapso esférico para formação de halos. Esse modelo foi proposto por \citet{PS1974} (PS) e foi amplamente utilizado até a década de 1990, quando simulações numéricas de N-corpos começaram a ser utilizadas para obter parametrizações mais acuradas para a função de massa. Por exemplo, \citet{Sheth1999} consideraram o colapso elíptico, enquanto \citet{Tinker2008} levam em conta a dependência em relação ao redshift e sobredensidade em relação à densidade de referência  (posteriormente outros autores fazem essas considerações). \citet{Watson2013} ajustam e comparam as funções obtidas com dois métodos distintos de identificação de halos (\textit{halo finder})~\citep{Knebe2013} - o \textit{Friends of Friends (FoF)}~\citep{Huchra1982, Press1982, Davis1985} e o \textit{Spherical Overdensity} (SO). \cite{Bocquet2016} investigam o impacto dos bárions.\footnote{Há várias outras funções multiplicidade disponíveis na literatura. A implementação de muitas delas estão disponíveis nos pacotes \textit{Numerical Cosmology Library} (NumCosmo)~\citep{NumCosmo2014} e \textit{Core Cosmology Library} (CCL)~\citep{CCL2019}. \\ \url{https://github.com/NumCosmo/NumCosmo} \\ \url{https://github.com/LSSTDESC/CCL}}

\section{Indicadores de massa}
\label{sec:proxy}

Até o momento apresentamos a contagem de halos de matéria escura, contudo, na prática, queremos obter a abundância de aglomerados de galáxias e, portanto, precisamos definir como conectar essas duas quantidades. Primeiramente, considerando massas típicas de grupos e aglomerados de galáxias, temos que cada halo está associado a um aglomerado. As Eqs.~\eqref{eq:d2N} e \eqref{eq:mass_function} dependem da massa e redshift dos objetos, ou seja, $M$ e $z$ dos halos/aglomerados. No entanto, as massas dos aglomerados não podem ser medidas diretamente e, portanto, precisamos identificar traçadores de massas (\textit{mass proxies}) para obter estimativas de $M$. 

Como mencionamos na Sec.~\ref{sec:introducao}, os aglomerados são compostos por matéria escura, ICM e galáxias. Cada uma dessas componentes fornece sinais em diferentes bandas do espectro eletromagnético (ou sinal ``gravitacional'' no caso da matéria escura). Veremos a seguir a relação entre esses sinais e as respectivas componentes dos aglomerados.

\subsection{Raio-X}

As observações em raio-X traçam informações de grande parte da matéria bariônica dos aglomerados. Em particular, o ICM emite radiação Raio-X oriunda da colisão entre elétrons livres e íons devido ao movimento térmico aleatório dessas partículas (Bremsstrahlung). Os traçadores de massa são a luminosidade em raio-X $L_X$, temperatura do gás $T_X$ e a energia térmica $Y_X = M_{\text{gás}} kT_X$, onde $M_{\text{gás}}$ é a massa do ICM e $k$ é a constante de Boltzmann~\citep{Rosati2002}. A relação entre $L_X - T_x$, em geral, supõe equilíbrio hidrostático. Consequentemente, essa condição pode fornecer uma relação massa-observável com uma calibração ``espúria'' caso o catálogo de aglomerados selecionado contenha muitos objetos que estejam longe do equilíbrio. Algumas características dos aglomerados, como \textit{cool core} e o estado dinâmico, são por vezes levadas em conta para se obter resultados acurados~\citep{Giles2016, Lieu2016}. 

\subsection{Milimétrico} 

As observações (identificação) de aglomerados na faixa milimétrica (mm) do espectro ocorre devido ao espalhamento Compton inverso dos fótons da radiação cósmica de fundo com os elétrons do ICM. Esse é o chamado Efeito Sunyaev–Zeldovich (SZ)~\citep{SZ1972}. O principal efeito SZ é o térmico e a partir dele é medido o parâmetro de Compton $Y_{SZ}$, o qual depende da temperatura e densidade do gás (ICM). A superfície de brilho do efeito SZ é independente do redshit do aglomerado. Essa é uma grande vantagem para identificar aglomerados em altos redshift ($1 \lesssim z \lesssim 2$), e também para estudar a evolução das relações de escala (massa-observável)~\citep{Bocquet2019}.

\subsection{Visível e infravermelho próximo}

A emissão estelar das galáxias e luz do ICM são os sinais emitidos no visível e infravermelho próximo. Medidas em bandas do infravermelho são apropriadas para identificar aglomerados em mais altos redshift. Nos levantamentos óticos, utiliza-se a riqueza dos aglomerados como traçador de suas massas. Existem várias formas de se definir riqueza $\lambda$. Uma delas corresponde ao número de galáxias elípticas vermelhas presentes no aglomerado. Essa aparente liberdade na definição de $\lambda$ ocorre pois a mesma depende da maneira de se identificar os aglomerados nas imagens, ou seja, depende dos \textit{(optical) cluster finders}. Citamos aqui alguns desses algoritmos: \textit{Adaptive Matched Identifier of Clustered Objects} (AMICO)~\citep{Bellagamba2018}, \textit{red-sequence Matched-filter Probabilistic
Percolation } (redMaPPer)~\citep{Rykoff2014}, \textit{Wavelet Adapted Z Photometric} (WaZP)~\citep{Dietrich2014, Aguena2021}, entre outros. 
 
Além da riqueza, os levantamentos óticos também podem fornecer outras medidas, referentes aos efeitos de lenteamento gravitacional fraco e forte, para estimarmos as massas dos aglomerados. O efeito de lenteamento gravitacional ocorre devido à presença de um objeto massivo (e.g., galáxia ou aglomerado de galáxia) que gera um campo gravitacional e, portanto, distorce o espaço-tempo na região onde o mesmo se encontra e na sua vizinhança. Essa distorção (que faz com que o objeto atue como uma lente) altera as trajetórias dos fótons emitidos por objetos fontes\footnote{Consideramos um observador, fontes e uma lente que encontra-se entre eles.}. No caso do efeito forte, observam-se imagens múltiplas de objetos que são lenteados pelo aglomerado e/ou arcos gravitacionais. O sinal de lenteamento gravitacional fraco gera uma distorção na elipticidade das galáxias e, também, pode intensificar o brilho das mesmas (magnificação). Veja~\citet{Umetsu2020} para mais detalhes, e \citet{Zitrin2015} como exemplo de método para estimar massa de aglomerados combinando medidas de lenteamento gravitacional forte e fraco.

Diferente dos indicadores de massa mencionados acima (e.g., $Y_X$, $Y_{SZ}$ e $\lambda$), que estão relacionados à basicamente uma componente do aglomerado (ICM ou galáxias), os sinais de lenteamento gravitacional traçam a massa total do aglomerado (matéria escura + bariônica). Além disso, o lenteamento gravitacional não é sensível à astrofísica do gás. Por isso, o lenteamento gravitacional é considerado o melhor candidato para determinar a calibração de massa absoluta.\footnote{Neste sentido, consideramos que os outros traçadores podem fornecer somente uma calibração de massa relativa, já que eles não rastreiam a massa total do aglomerado.}  


\section{Verossimilhanças}
\label{sec:likelihood}

Dada uma relação massa-observável, podemos então determinar a densidade número de aglomerados de galáxias. Digamos, por exemplo, que temos um conjunto de medidas de riqueza e redshift fotométrico $\{\lambda_i, z^{\text{phot}}_i\}$, onde $i$ representa o índice do i-ésimo aglomerado. Essas medidas estão associadas às variáveis aleatórias $\lambda$ e $z$, que, por conveniência, chamaremos de $\lambda^{\text{real}}$ e $z^{\text{real}}$. Portanto, temos que as quantidades medidas serão descritas por distribuições de probabilidade das quantidades ``reais". Isto é, $P(\lambda_i | \lambda^{\text{real}}) \ \rm{d} \lambda^{\text{real}}$ é a distribuição de probabilidade de medir $\lambda_i$ dado $\lambda^{\text{real}}$. Analogamente, consideramos também as distribuições de probabilidade do redshift $P(z^{\text{phot}}_i | z^{\text{real}})$, e da relação de escala $P(\lambda^{\text{real}} | \ln M)$. Dessa forma, o número médio de aglomerados de galáxias com redshift no intervalo $[z^{\text{phot}}, z^{\text{phot}} + dz^{\text{phot}}]$ e riqueza $[\lambda, \lambda + d\lambda]$ é 

\begin{eqnarray}
		\label{eq:d2N_z:xi}
		\frac{\rm{d}^2N(\lambda_i, z^{\text{phot}}_i, \vec{\theta})}{\rm{d}z^{\text{phot}} \rm{d}\lambda} &=& \int \rm{d}\ln M \int \rm{d}\lambda^{\text{real}} \int \rm{d}z^{\text{real}} \, \Phi(M,z)  \\
		&\times & \frac{\rm{d}^2N(M, z^{\text{real}}, \vec{\theta})}{\rm{d}z^{\text{real}} \rm{d}\ln M} {P(\lambda_i | \lambda^{\text{real}}) \, P(\lambda^{\text{real}} | \ln M, z^{\text{real}})} \, P(z^{\text{phot}}_i | z^{\text{real}}), \nonumber
\end{eqnarray}
onde $\vec{\theta}$ representa o conjunto de parâmetros da modelagem (e.g., cosmológicos, e da relação massa-observável), e $\Phi(M,z)$ é a função seleção que descreve a pureza e a completeza do catálogo de aglomerados. Como mencionamos anteriormente, em princípio temos uma relação de um para um entre halos de matéria escura e aglomerados de galáxias. No entanto, os \textit{cluster finders} possuem limitações e podem tanto falhar em identificar alguns aglomerados e, com isso, o catálogo produto será incompleto, quanto podem classificar erroneamente um outro tipo objeto como sendo um aglomerado, ou seja, o catálogo final será impuro. Em suma, completeza é a razão entre o número de aglomerados detectados e o número real de halos de matéria escura, e a pureza é a razão entre o número de aglomerados corretamente identificados e o número total de aglomerados detectados~\citep{Aguena2018}.

Finalmente, a partir da Eq.~\eqref{eq:d2N_z:xi}, podemos construir uma verossimilhança (probabilidade de medir um certo conjunto de dados, amostra, como função de parâmetros de modelo estatístico) para realizar análises cosmológicas e/ou calibrar a relação massa-observável. Uma verossimilhança da abundância de aglomerados possui, por construção,  um erro Poissoniano. Além disso, pela teoria de formação de estruturas, temos que halos (aglomerados, galáxias) são traçadores enviesados do campo de densidade de matéria~\citep{Cooray2002}. Portanto, temos que o número médio de aglomerados também está sujeito à distribuição espacial da matéria, ou seja, à estrutura em grande escala do universo. Isso gera o chamado \textit{(super-) sample variance}~\citep{Hu2003, Takada2013, Takada2014}.

Em geral, nos levantamentos de raio-X e SZ (mm), o número de aglomerados nos catálogos e as suas massas são tais que o erro de \textit{sample variance} é desprezível e, portanto, pode-se considerar somente o erro de Poisson (\textit{shot-noise}). Nesse contexto, é muito utilizada a verossimilhança de contagem de aglomerados não binada~\citep{Penna2014, Planck2015, Bocquet2019} (em geral a binagem é feita no traçador de massa e no redshift).

Por sua vez, os levantamentos óticos produzem (mais comumente) catálogos com um número muito maior de aglomerados atingindo massa mais baixas $\sim 5 \times 10^{13} \, M_\odot$. Por isso, o termo de \textit{sample variance} tem que ser levado em conta nas análises. Neste caso, geralmente constrói-se uma verossimilhança para abundância de aglomerados Gaussiana binada na riqueza e no redshift~\citep{Constanzi2019}.  

Introduzimos uma terceira verossimilhança, que chamamos de \textit{pseudo-counts}, para ser aplicada em situações onde não temos conhecimento (ou uma boa determinação) da função seleção~\citep{Penna2017}. A função seleção também possui uma incerteza associada. Aplicamos esse método para calibrar relações de escala de lenteamento e SZ, levando em conta uma correlação entre elas. A grande vantagem do \textit{pseudo-counts} é poder realizar uma análise completa de calibração de massa, cosmológica e função de seleção.



	

\section{Fontes de incerteza}
\label{sec:incerteza}

A partir das definições e discussões que realizamos, vamos agora exemplificar algumas fontes de incerteza~\footnote{Nesta seção, não mencionaremos as incertezas oriundas dos \textit{cluster finders}, como a função seleção que discutimos na Sec.~\ref{sec:likelihood} ou os efeitos de projeção~\citep{Sunayama2020}, entre outros.} e requerimentos para levantamentos como o LSST.

\subsection{Função multiplicidade}

As funções multiplicidade têm sido obtidas a partir de simulações de N-corpos, as quais cada vez mais incluem efeitos e características mais realistas. No entanto, precisamos lembrar que essas funções dependem da definição de massa, dos algoritmos de \textit{halo finders}, do tipo de simulação, cosmologia ou uma suíte de cosmologias (mesmo que fracamente), etc., e esses fatores têm de ser levados em conta coerentemente nas análises. Para análises e discussões mais extensas, vejam~\citet{Knebe2013} e \citet{Artis2021} (e referências contidas nesses trabalhos).

Na prática, para caracterizar de forma mais concreta, temos que a função multiplicidade de \citet{Tinker2008} possui precisão da ordem de 5\% em $z = 0$. Já o emulador desenvolvido por \citet{McClintock2019} fornece uma função com 1\% precisão em $z = 0$. E é esse nível de precisão que será requerido pelo LSST, por exemplo.

\subsection{Estimadores dos parâmetros cosmológicos}

Em \citep{Penna2014}, nós estudamos os estimadores dos parâmetros cosmológicos $\Omega_c$ (densidade de matéria escura fria), $\sigma_8$ (desvio padrão do contraste de densidade de matéria na escala $R = 8$ h${}^{-1}$ Mpc) e $w$ (equação de estado da energia escura, supondo que ela seja constante) utilizando uma verossimilhança sem binagem da abundância de aglomerados de galáxias. As análises foram feitas em diferentes cenários a fim de considerarmos as influências dos seguintes elementos no viés dos parâmetros:
\begin{itemize}
    \item redshift máximo do levantamento;
    \item redshift espectroscópico e redshift fotométrico;
    \item área angular do levantamento;
    \item número de parâmetros ajustados simultaneamente;
    \item conhecimento da massa real;
    \item incerteza na massa.
\end{itemize}

Verificamos que, em geral, os estimadores desses parâmetros são enviesados, mas esses vieses são pequenos quando comparados aos respectivos erros. No entanto, encontramos alguns casos (grande área angular), em que os vieses podem ser relevantes, da ordem do tamanho da barra de erro e, portanto, essa investigação deve ser feita caso a caso. Combinando a verossimilhança sem binagem da abundância de aglomerados de galáxias com uma verossimilhança da radiação cósmica de fundo, obtivemos estimadores de $\Omega_c$, $\sigma_8$ e $w$ sem vieses. Finalmente, quando consideramos verossimilhanças binadas da contagem de aglomerados, obtivemos indicações de que os vieses podem ser ainda maiores do que no caso de verossimilhanças não binadas~\citep{PennaLima2012}.

\subsection{Redshift fotométrico}

Vimos na Sec.~\ref{sec:contagem} que o redshift dos halos/aglomerados de galáxias é uma das quantidades que precisamos medir para obter a contagem desses objetos. O redshift pode ser medido por espectroscopia ou por fotometria. Essa última produz resultados com maiores barras de erro, mas os catálogos possuem, em geral, um número muito maior de objetos. Muitos levantamentos utilizam este método e, portanto, precisamos modelar o erro dessas medidas para calcular a abundância de aglomerados (veja a Eq.~\eqref{eq:d2N_z:xi}). 

A distribuição de probabilidade $P(z^{\text{phot}}_i | z^{\text{real}})$ pode ser obtida empiricamente dos dados e incluídas uma a uma na verossimilhança não binada. Uma aproximação muito utilizada é considerar que o redshift fotométrico, $z^{\text{phot}} = z^{\text{true}} + z^{\text{bias}} \pm \sigma_z$, segue uma distribuição Gaussiana da seguinte forma,  
\begin{equation}
    P(z^{\text{phot}} | z^{\text{true}}) = \sqrt{\frac{2}{\pi}} \frac{e^{-\frac{\left( z^{\text{phot}} -  z^{\text{true}} \right)^2}{2\sigma_z^2}}}{\sigma_z \left( 1 - \text{erf} \left(z^{\text{true}} / \sqrt{2\sigma_z^2}\right) \right)},
\end{equation}
onde $\sigma_z = \sigma_z^0 (1+z)$.

\begin{itemize}		
		\item Grandes Levantamentos: 
		\begin{itemize}
		    \item \textit{Dark Energy Survey} (DES)~\citep{DES2005, Abbott2020} -- 5 filtros, 5.000 deg$^2$, $\sigma_z^0 = 0,03$, $z \lesssim 1.4$;
		    \item \textit{Javalambre Physics of the Accelerating Universe Astrophysical Survey} (J-PAS) -- 56 filtros, 8.500 deg$^2$, $\sigma_z^0 = 0,003$, $z \lesssim 1,5$;
		    \item Satélite Euclid -- 7 filtros, $\sigma_z^0 = 0,025$ -- 0,053, $z \lesssim 2,0$;
		    \item LSST -- 6 filtros (ugrizy), 18.000 deg$^2$, $\sigma_z^0 \leq 0,02$, $z^{\text{bias}} < 0,003$, $z \lesssim 1,2$.
		\end{itemize}
	\end{itemize}

\subsection{Relações de escala (massa-observável)}

Para finalizar, vamos dar continuidade à questão da estimar as massas dos aglomerados. Como mencionamos anteriormente, a massa não pode ser observada diretamente e é a principal fonte de incerteza nos estudos cosmológicos com aglomerados de galáxias. Em princípio, conseguiremos acessar todo o potencial dos aglomerados de galáxias como sonda cosmológica, caso as relações de escala (massa-observável) sejam calibradas nos intervalos de massa e redshift dentro de um nível de dispersão de 5\%~\citep{Wu2010}. Listamos abaixo algumas questões relacionadas à calibração de massa.

\begin{itemize}
\item Traçadores em raio-X: relação $L_X - M$ com dispersão da ordem de 40\%, $T_X - M$ dispersão $\sim 10 - 15\%$~\citep{Mantz2016}. 
\item Lenteamento gravitacional fraco é o traçador de massa mais promissor, já que ele traça a massa absoluta e não é sensível à astrofísica do gás;
\item Estimativas de massa individuais via lenteamento fraco apresentam vieses menores do que os de raio-X, mas possuem mais ruído;
\item Lenteamento fraco requer a calibração das medidas de cisalhamento;
\item É muito importante utilizar dados combinados em múltiplos comprimentos de onda para medir relações massa-observável com menores dispersões.
\end{itemize}

\section{Considerações finais}
\label{sec:conclusao}

Neste trabalho, nós fizemos uma breve revisão sobre o uso da contagem de aglomerados de galáxias como sonda cosmológica, dando ênfase a diferentes fontes de incerteza que devem ser tratadas cuidadosamente, principalmente para os próximos levantamentos como o LSST e o satélite Euclid, para que possamos acessar o máximo potencial dos aglomerados em restringir alguns parâmetros cosmológicos, e.g., densidade de matéria $\Omega_m$, o desvio padrão do contraste de densidade de matéria filtrado na escala $R = 8 \ \text{h}^{-1}$ Mpc $\sigma_8$, e a constante de Hubble $H_0$.
Além disso, o desenvolvimento e melhoria desses trabalhos e análises não contribuem somente com os estudos cosmológicos, mas também no entendimento da dinâmica, formação e evolução dos aglomerados de galáxias.

A distribuição espacial dos aglomerados de galáxias (\textit{cluster clustering}) também é uma sonda cosmológica muito importante. A sua descrição utiliza os mesmos elementos necessários para a contagem, mas também o viés do halo (\textit{halo bias}) e o perfil de densidade de matéria do halo. Assim como a função multiplicidade, por exemplo, esses termos também são extensamente estudados e ajustados a partir de simulações de N-corpos. Muitos dos perfis de densidade de matéria supõem simétrica esférica~\citep{Navarro1995, Einasto1965}, o que não descreve adequadamente muitos dos aglomerados (principalmente aqueles que estão longe do equilíbrio hidrodinâmico). Com isso, correções como inclusão de triaxialidade, podem ser feitas e exploradas a fim de fornecer uma descrição mais realista para o perfil de densidade dos aglomerados~\citep{Herbonnet2021}. Vale enfatizar que o \textit{halo bias} está presente nos estudos com contagem de aglomerados ((super-) \textit{sample variance}), e o perfil de densidade de matéria está presente nas análises de lenteamento gravitacional como traçador da massa dos aglomerados.

Finalmente, um último ponto extremamente importante é a qualidade dos códigos numéricos. Como estamos falando de restrição de parâmetros cosmológicos com barras de erro de poucos porcentos, é necessário que os códigos numéricos realizem os cálculos com precisão cada vez maiores. Não é mais viável calcular quantidades com precisão relativa de $10^{-3} - 10^{-4}$. A princípio, isso pode parecer ser um problema bem simples. No entanto, o custo computacional para aumentar as precisões dos cálculos é, em geral, muito grande. Além da precisão, todo o refinamento da modelagem astrofísica dos observáveis traz complexidades numéricas que requerem muito esforço na questão de otimização dos códigos sem comprometer a precisão do cálculo. Veja~\cite{Vitenti2019} para uma discussão extensa sobre métodos numéricos.


\section*{Agradecimentos}

Meu agradecimento aos(às) organizadores(as) do evento \textit{I Encontro Brasileiro de Meninas e Mulheres da Astrofísica, Cosmologia e Gravitação}. Agradeço também a Eduardo José Barroso e a Sandro Vitenti pelos comentários e sugestões ao texto.


\bibliographystyle{unsrtnat}
\bibliography{references}  



\end{document}